\RequirePackage[2020-02-02]{latexrelease}

\documentclass{iau}

\usepackage{amsmath}
\usepackage{graphicx}
\usepackage{multirow}
\usepackage{hyperref}

\lefttitle{E. Kasai et. al}
\righttitle{Redshift determination of blazars for CTA}

\doival{}

\aopheadtitle{Proceedings of the IAU Symposium 375}
\editors{Y. Liodakis et al. 2023}

\title{Redshift determination of blazars for the Cherenkov Telescope Array}

\author{E. Kasai$^{1}$, P. Goldoni$^{2}$, S. Pita$^{3}$, C. Boisson$^{4}$, M. Backes$^{1,5}$, G. Cotter$^{6}$, F.~D'Ammando$^{7}$, B. van Soelen$^{8}$}
\affiliation{$^{1}$Department of Physics, Chemistry \& Material Science, University of Namibia, Private Bag 13301, Windhoek, Namibia \\
$^{2}$Universit\'e Paris Cit\'e, CNRS, CEA, Astroparticule et Cosmologie, F-75013 Paris, France \\
$^{3}$Universit\'e Paris Cit\'e, CNRS, Astroparticule et Cosmologie, F-75013 Paris, France \\
$^{4}$LUTH, Observatoire de Paris,  PSL Research University, CNRS, Universit\'{e} Paris Cit\'e, Meudon, France \\
$^{5}$Centre for Space Research, North-West University, Potchefstroom 2520, South Africa \\
$^{6}$University of Oxford, Oxford Astrophysics, Denys Wilkinson Building, Keble Road, Oxford, OX1 3RH, United Kingdom \\
$^{7}$INAF - Istituto di Radioastronomia, Via Gobetti 101, I-40129 Bologna, Italy \\
$^{8}$Department of Physics, University of the Free State, Bloemfontein 9300, South Africa}

\begin{document}

\begin{abstract}
Blazars are the most numerous type of observed high-energy gamma-ray emitters. However, their emission mechanisms and population properties are still not well-understood. Crucial to this understanding are their cosmological redshifts, which are often not easy to obtain. This presents a great challenge to the next-generation ground-based observatory for very-high-energy gamma rays, the Cherenkov Telescope Array (CTA), which aims to detect a large number of distant blazars to study their intrinsic emission properties and to place tight constraints on the extragalactic background light density, amongst others. The successful investigation of these subjects needs a precise redshift determination. Motivated by these challenges, the CTA redshift task force initiated more than 3 years ago a spectroscopic observing program using some of the largest optical and infrared telescopes to measure the redshifts of a large fraction of blazars that are likely to be detected with CTA. In this proceedings, we give an overview of the CTA redshift task force, discuss some of the difficulties associated with measuring the redshifts of blazars and present our sample selection and observing strategies. We end the proceedings with reporting selected results from the program, the on-going collaborative efforts and our plans for the future.
\end{abstract}


\maketitle

\section{Introduction}
\label{intro}
Blazars are a type of active galactic nuclei (AGNs) characterized by highly powered relativistic jets that are directed nearly along the line of sight of an observer. They are divided into flat spectrum radio quasars (FSRQs) and BL Lacertae objects (BL Lacs). In the optical regime, FSRQs are characterized by strong and broad emission lines whereas BL Lacs are continuum-dominated with their spectral features having equivalent widths (EWs) of less than 5\AA~\citep{1995PASP..107..803U}. The very-high-energy emission from blazars enables the study of the extragalactic background light density and investigations of the intergalactic magnetic field, Lorentz invariance violation as well as theories of the existence of axion-like particles \citep[][and references therein]{2019scta.book.....C}. The successful investigation of each of these topics heavily relies on blazar redshift measurements, an exercise that is quite difficult to conduct. In fact, excluding FSRQs and blazar candidates of unknown type, 1897 of 2951 {\it Fermi}-LAT detected blazars have no redshifts. It is thus of utmost importance that efforts are made to measure as many redshifts as possible for general blazar science with CTA. For this reason, a task force was formed within the CTA Consortium more than three years ago, to measure the redshifts of many blazars across cosmic time. 

This proceedings is organized as follows: we give an overview of the CTA redshift task force in Section 2, followed by a discussion of the difficulties we face in measuring the redshifts of blazars in Section 3. We describe our sample selection process and observing strategy in Section 4 and present selected results in Section 5. In Section 6, we provide a short summary and briefly discuss our plans for the future.

\section{The CTA redshift task force and its objectives}
The CTA redshift task force (hereafter the task force) is a CTA working group primarily created with the goal to measure the redshifts of blazars using some of the largest optical and infrared telescopes to support the CTA Key Science Project on AGNs of blazar type \citep{2019scta.book.....C}. Table~\autoref{tab1} provides a list of some of the spectroscopic instruments and telescopes used by the task force. In the last column of the table are the combined total numbers of sources observed by each telescope, published in the first two papers by the task force \citep{2021A&A...650A.106G, 2023MNRAS.518.2675K} and are {\it to be published in the upcoming third paper}.

\begin{table}[h!]
\caption{\label{tab1} Instrument name, observatory name, diameter of primary mirror, wavelength coverage, spectral resolution and combined total number of sources for which spectra have been published in the first two papers by the task force and are to be published in the upcoming third paper.}

\centering
\begin{tabular}{lcccccc}
\hline\hline

Instrument & Observatory & Mirror&  Wavelength  & Spectral & Total \\
name & name  & size (m) & coverage (\AA) & resolution ($\lambda$/$\Delta\lambda$) & sources \\
\hline

Keck/ESI  & W. M. Keck & 10 & 3900 -- 10000 & $\sim$ 10000 & 14\\
SALT/RSS & Sutherland & 11 & 4500 -- 7500 & $\sim$ 1000 & 15 \\
NTT/EFOSC2 & La Silla & 3.5 & 3860 -- 8070 & $\sim$ 500 & 18 \\
Lick/KAST & Lick & 3 & $\sim$ 3500 -- 5600 & $\sim$ 1000 & \multirow{2}{*} {37} \\ \cline{7-7}
 & &  & $\sim$ 5400 -- 8000 & $\sim$ 1500 &  \\%\multirow{2}{*}{} \\ \cline{6-6}
VLT/FORS2 & Paranal & 8.2 & 3300 -- 11000 & 260 -- 1600 & 14 \\

\hline
\end{tabular}
\end{table}

%Further details of the instruments in Table \ref{tab1} and the telescopes they are mounted on can be accessed from the websites of the provided observatory names.
Further details of the instruments in Table \autoref{tab1} are accessible via the websites of the observatories that are home to the telescopes on which they are mounted.

\section {Blazar redshift measurement difficulties}
The challenging aspect in measuring blazar redshifts stems from the overwhelming non-thermal emission of the jet that, for the most part, outshines the host galaxy thermal emission. While this is not the case with FSRQs that are characterized by strong emission lines, it is so in the case of BL Lacs, whose spectra are characterised by weak host galaxy spectral features that are often quite difficult to detect, even with the best optical or infrared telescopes. The probability of a successful redshift measurement is thus difficult to estimate a-priori. 

Because of such challenges, the results of this type of observations are often difficult to interpret. This leads to conflicting conclusions among the various groups of researchers involved in this type of work. Compiling a detailed report on each observed source, including classification details, evidence of extension and previous observational results, is therefore a crucial undertaking. In our attempt to get around such difficulties and still be able to measure the redshift of a given blazar or obtain a lower limit of it, we apply several strategies. These include (1) obtaining high S/N ($\ge$ 100) spectra, (2) searching for absorption systems along the line of sight, (3) detection of host galaxies in deep imaging observations and (4) conducting spectroscopic observations during periods of low optical blazar activity.

\section{Sample selection and observing strategy}
Our sources were selected from the third {\it Fermi}-LAT catalog of high energy sources -- abbreviated 3FHL \citep{2017ApJS..232...18A} -- at the onset of the project more than three years ago. The catalogue had 1040 BL Lac candidates and BCUs (blazar candidates of unknown type) without redshifts. The selection process was performed with the help of Monte Carlo (MC) simulations using the Gammapy software\footnote{\url{https://gammapy.org}} to estimate the minimum CTA observation time for each source that yields a 5$\sigma$ detection. Exact details of the process are presented in \citet{2021A&A...650A.106G}. The sample selection process resulted in the selection of 165 sources that CTA is able to detect in under 30 hours at 5$\sigma$.

\begin{figure}
  \centering
    \includegraphics[scale=.45]{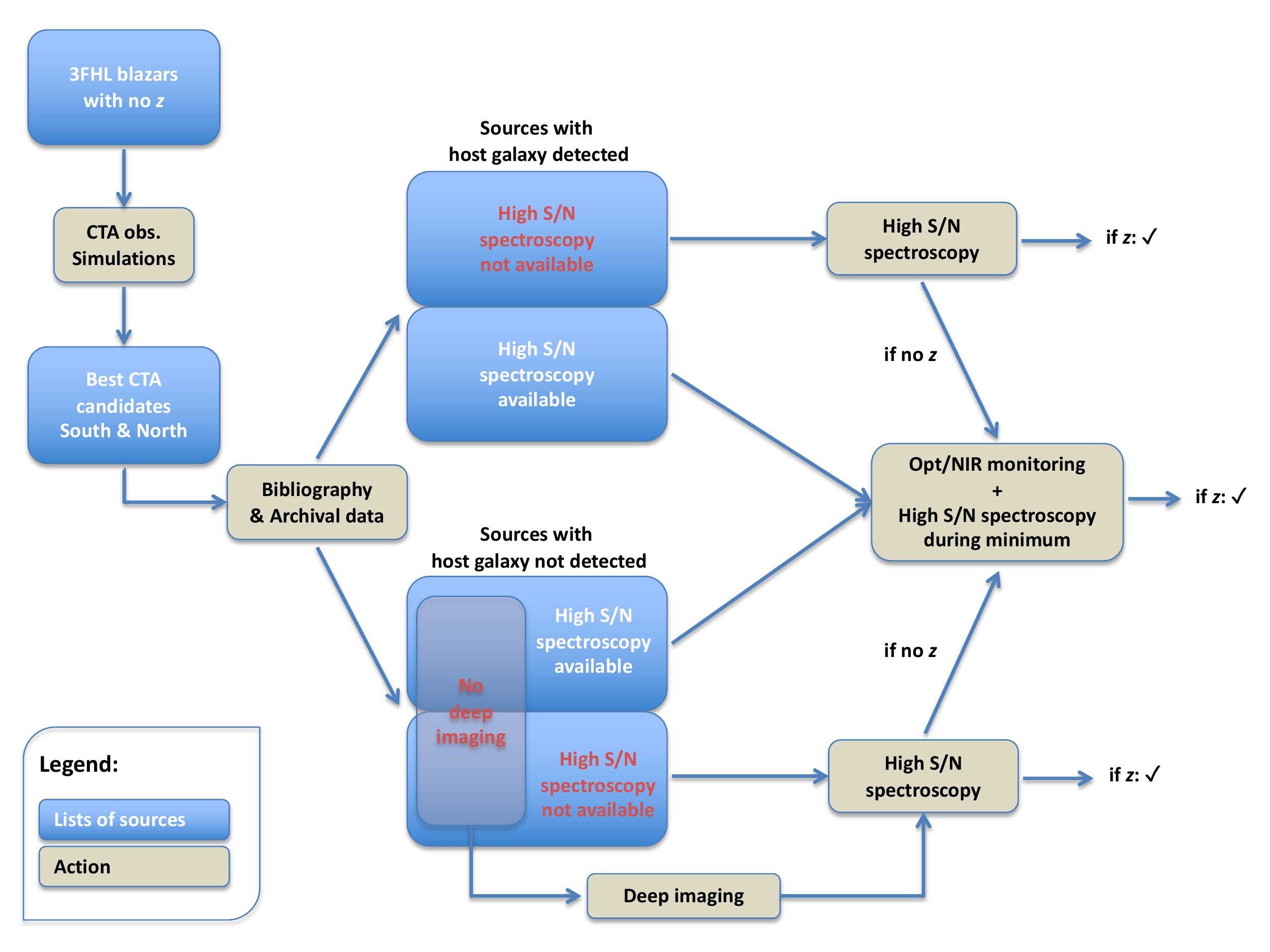}
    \caption{A flow-diagram summarizing the CTA redshift task force source sample selection process and observing strategy.}
    \label{figure1}
  \end{figure}

Our observing strategy is centred on searching for host galaxy stellar absorption and emission features in the spectra. To achieve EW sensitivities of less than 5\AA~for both absorption and emission features, we require that each spectrum has (1) a spectral resolution $\lambda/\Delta \lambda$ of 1000 or higher and (2) an average S/N ratio of 100 or higher per pixel. For each source, we look for previous spectroscopic results and evidence of extension in the literature, e.g. the 2MASX catalogue \citep{2000AJ....119.2498J}. This leads to the classification of sources as high-priority or low-priority. Figure \autoref{figure1} shows a flow-diagram summarizing our sample selection process and observing strategy. 

\section{Selected results}
\begin{figure}[h!]
   \centering
\includegraphics[width=6.5truecm,height=5.1truecm]{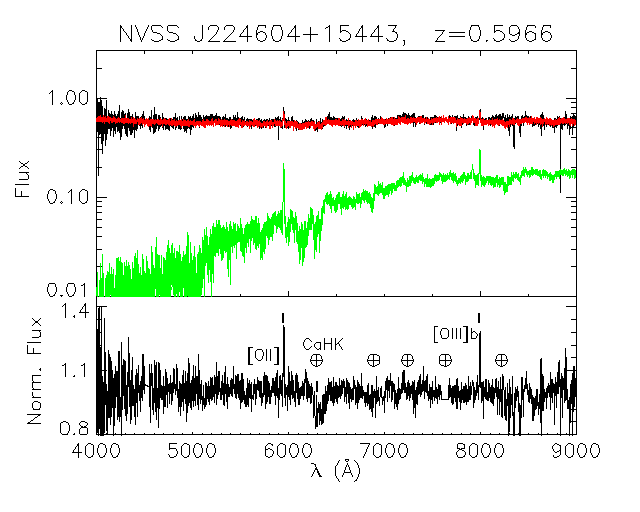}\includegraphics[width=6.5truecm,height=5.1truecm]{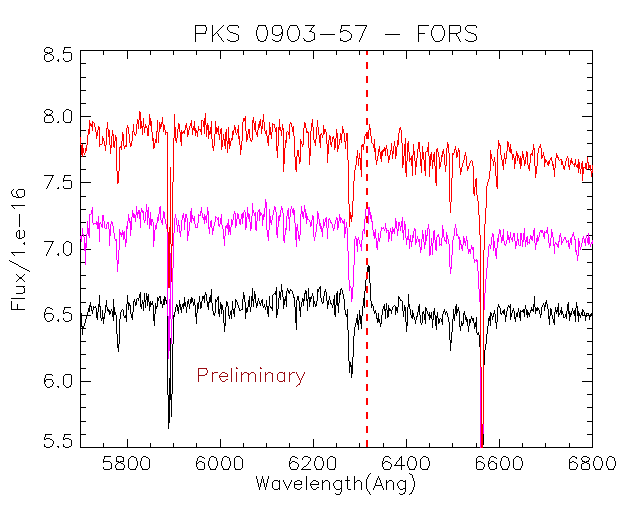}

\caption{{\it Left}: Observational results for the source NVSS 224604+15443 published in the second paper (Figure 5, Optical Spectroscopy of Blazars for the Cherenkov Telescope Array -- II, Kasai {\it et al.} 2023, MNRAS, 518). In the {\it upper} panel are the flux-calibrated and telluric-corrected spectrum (black) and jet+galaxy best-fitting model (red). The elliptical galaxy component of the model is shown in green. The {\it lower} panel contains the normalised spectrum. The detected spectral features are labelled accordingly. The symbol $\oplus$ represents atmospheric telluric absorption features. {\it Right}: Three preliminary VLT/FORS2 spectra of the blazar PKS 0903-57, with an uncertain redshift. An emission feature, marked by the dashed red line, is visible. The strong absorption features around 5890 \AA~and 6560 \AA~are, respectively, due to the Milky Way ISM and to a nearby star which lies at 0.67 arcseconds and whose spectrum is superposed onto the one for the blazar.}
\label{fig2}
\end{figure}

We present selected results from the two papers published by the task force and preliminary VLT/FORS2 spectra of the blazar PKS0903-57 with an uncertain redshift. \citet{2021A&A...650A.106G} presented spectra for 19 sources, eleven of which had their redshifts determined successfully. \citet{2023MNRAS.518.2675K} presented spectra for 33 sources (including 13 spectra for eight sources that were of lower quality and featureless). Fourteen of the sources had their redshifts determined successfully. The left side of Figure \autoref{fig2} shows observational results for one source published in the second paper and for which a redshift (given in the subfigure title) was measured. The right side of the figure shows three preliminary VLT/FORS2 spectra of the blazar PKS 0903-57.

At the time of this proceedings, the analysis of the three spectra to determine the significance level of the detected emission feature -- marked with the vertical dashed line in the figure -- has not been finalized. Further observations are needed to determine the redshift of PKS~0903-57.

\section{Summary and future work}
To summarize, the task force is a CTA working group tasked with measuring redshifts of blazars that are likely to be detected by CTA and seen as an important support arm for the CTA Key Science Project on AGNs of blazar type. We continue spectroscopic and photometric observations of our sources, and the publication of our results. Our first imaging paper, third spectroscopy paper and a public web database of our observations are currently in preparation.

%\begin{thebibliography}
%\bibitem[Urry(1995)]{1995PASP..107..803U} Urry, C.\ 1995, 
%PASP, 107, 803
%\bibitem[Collier Cameron(1999)]{1999ASPC..158..146C} Collier Cameron, A.\ 1999, Solar and Stellar Activity: Similarities and Differences, 146
%\bibitem[Donati~{\it et. al}(1992)]{1992A&A...265..682D} Donati, J.-F., Brown, S.~F., Semel, M., {\it et. al}\ 1992, {\it A\&A}, 265, 682
%\end{thebibliography}

\end{document}